\documentstyle[prl,aps,twocolumn]{revtex}
\begin{document}
\draft
\flushbottom
\twocolumn[\hsize\textwidth\columnwidth\hsize\csname
@twocolumnfalse\endcsname

\title{\bf Melting of sodium clusters}

\author{
 Juan A. Reyes-Nava$^{1}$,
 Ignacio L. Garz\'on$^{1}$,
 Marcela R.  Beltr\'an$^{2}$, and
 Karo Michaelian$^{1}$
}

\address{
$^1$Instituto de F\'{\i}sica,
    Universidad Nacional Aut\'onoma de M\'exico,
    Apartado Postal 20-364, M\'exico D.F., 01000 M\'exico \\
$^2$Instituto de Investigaciones en Materiales,
    Universidad Nacional Aut\'onoma de M\'exico, \\
    Apartado Postal 70-360, M\'exico D.F., 01000 M\'exico \\
}

\maketitle

\begin{abstract}
Thermal stability properties  and the melting-like
transition of Na$_N$, $N$ = 13-147, 
clusters are studied through microcanonical molecular dynamics
simulations. The metallic bonding in the sodium clusters is mimicked  
by a many-body Gupta potential based on the second
moment approximation of a tight-binding Hamiltonian.
The characteristics of the solid-to-liquid transition in the sodium
clusters are analyzed by calculating 
physical quantities like caloric curves, heat capacities, and 
root-mean-square bond
length fluctuations using simulation times of several
nanoseconds. Distinct melting mechanisms are obtained for the 
sodium clusters in the size range investigated.
The calculated melting temperatures show an irregular variation
with the cluster size, in qualitative agreement with recent experimental 
results. However, the calculated melting point for the Na$_{55}$ cluster
is about 40 $\%$ lower than the experimental value.
\\
La fusi\'on y las propiedades de estabilidad t\'ermica de c\'umulos
de Na$_N$, $N$ = 13-147, se estudian utilizando simulaciones de
din\'amica molecular en el ensamble microcan\'onico. El enlace
met\'alico en los c\'umulos de sodio se modela con un potencial
de Gupta de muchos cuerpos que se basa en la aproximaci\'on de
segundo momento de un hamiltoniano de amarre fuerte. Las 
caracter\'{\i}sticas de la transici\'on s\'olido a l\'{\i}quido
en los c\'umulos de sodio se analizan mediante el c\'alculo
de cantidades f\'{\i}sicas como la curva cal\'orica, el calor
espec\'{\i}fico y la desviaci\'on cuadr\'atica media de las
fluctuaciones en las distancias interat\'omicas, utilizando tiempos
de simulaci\'on de varios nanosegundos. Mecanismos diferentes
de fusi\'on se obtuvieron para c\'umulos de sodio en el rango de
tama\~nos investigados. Las temperaturas de fusi\'on calculadas
muestran una variaci\'on irregular como funci\'on del tama\~no
del c\'umulo, en acuerdo cualitativo con resultados experimentales
recientes. Sin embargo, el punto de fusi\'on calculado para el
c\'umulo de Na$_{55}$ es aproximadamente 40 $\%$  m\'as bajo que el
valor experimental.\\
\\
\it Descriptores: \rm C\'umulos met\'alicos; c\'umulos de sodio;
fusi\'on en c\'umulos; transici\'on de fase en c\'umulos\\

\end{abstract}

]


\narrowtext

\section{Introduction}
The study of the  thermal stability and melting transition of
sodium clusters is 
nowadays a very active field of research. This interest is
motivated, in part, by the fact that just recently it was possible
to make direct comparisons between theoretical results, mainly extracted
from computer simulations 
\cite{And1,Bulgac1,Poteau,Manninen,Calvo1,Agua1,Calvo2,Agua2},
and experimental data \cite{Martin,Haber1,Haber2,Haber3} obtained for the
melting-like transition of sodium clusters of different sizes.
In fact, during the last few years the caloric curve, heat capacity,
and melting temperatures of Na$_N$, $N$ = 55-200, clusters
have been measured using the temperature dependence of the 
fragmentation mass spectra \cite{Haber1,Haber2,Haber3}.
The melting points of these clusters were found to be on
average 33 \% (120 K) lower than the bulk value, and more 
surprisingly, large variations were observed in the
melting temperatures ($\pm$ 30 K) with changing cluster
size \cite{Haber2}.

On the theoretical side, molecular dynamics (MD) and Monte Carlo
(MC) methods have been used to provide microscopic
descriptions on the melting mechanisms of sodium clusters in the
size range of 8-147 atoms. In these studies, the metallic bonding
of the sodium clusters has been described using different levels
of approximation, from empirical many-body model potentials
\cite{Bulgac1,Calvo1}
and tight-binding hamiltonians \cite{Poteau,Calvo2} 
to first-principles density functional theory 
\cite{And1,Manninen,Agua1,Agua2}.

Despite the large amount of information obtained from the above theoretical
studies, several questions on, for example, the irregular variations
of the melting temperatures with respect to the cluster size remain
unsolved. One difficulty existing in the theoretical approaches to study 
the melting-like transition of sodium clusters is related with the
interplay between the geometric and electronic structure effects
in such systems \cite{Calvo1,Agua1,Haber2}. Although MD simulations
that use density functional theory explicitly take into account
the electronic degrees of freedom and their effects in the
cluster geometry, the limitation of this
approach is related to the relatively short simulation times 
(a few picoseconds) during which time-averages of the structural
and dynamical properties are calculated. 
This problem, caused by the extremely large computational effort
involved in the first-principles MD simulations, is especially
critical in the study of the melting transition where large
thermal fluctuations are present, and therefore much longer simulation 
times are required to calculate converged time averages.

On the other hand, MD simulations based on many-body model potentials
allow extension of the simulation time up to the nanosecond regime,
employing reasonable computational resources. However, in this case
the description of the metallic bonding does not explicitly include
the electronic degrees of freedom. In the present work, we
adopt the latter approach to perform molecular dynamics
simulation of the melting-like transition of Na$_N$,
$N$ = 13, 20, 55, 135, 142, and 147, clusters using a many-body
Gupta potential \cite{Gupta,Rosato,Li}, and simulation times of  
$\sim $ 50 nanoseconds. The objectives of this study are: (1) to test the
phenomenological many-body Gupta potential in the description of the
melting transition using adequate simulation times and
(2) to compare the predictions of this simulation with
those obtained from the same many-body potential but using
the MC method \cite{Calvo1,Calvo2} for the averaging procedure, and also
with the results obtained from first-principles MD \cite{Agua1,Agua2} 
using shorter simulation times.
These tests and comparisons will provide additional
insights on the melting mechanisms in sodium clusters and
provide useful information on the performance of the
different simulation methods.

In Section II we describe the model potential and provide
details on the simulation method. Results on the caloric
curves, heat capacities and thermal stability properties of
different cluster sizes are given in Sec. III. A summary
of this work is presented in Sec. IV.

\section{Metallic potential and simulation procedure}

The many-body model potential used in this work corresponds to the
Gupta potential \cite{Gupta} that is based on the second
moment approximation of a tight-binding hamiltonian \cite{Rosato}.
Its analytical expression is given by:
\begin{equation}
\label{V}
V = \sum_{i=1}^{N} V_i
\end{equation}
\begin{equation}
V_i =  A \sum_{j \ne i} e^{-p \left(\frac{r_{ij}}{r_0}-1 \right)} -
  \xi \left( \sum_{j \ne i}
  e^{-2q \left(\frac{r_{ij}}{r_0}-1 \right)} \right)^{\frac{1}{2}}
\label{Gupta}
\end{equation}
where $r_0$, $A$,
$\xi$, $p$, and $q$ are adjustable parameters 
\cite{Rosato}. For sodium clusters these parameters have been
fitted to band structure calculations \cite{Li}. Their values
are: $A$=0.01595 eV, $\xi$=0.29113 eV, $r_0$=6.99 bohr, $p$=10.13,
and $q$=1.30 \cite{Li}.
This type of potential has been extensively utilized in other metal
cluster simulations \cite{Garzon1,Garzon2,Garzon3,Karo1,Karo2},
obtaining good agreement with results
generated from first-principles methods. 

In order to study the cluster melting-like transition,
we use the constant-energy MD method to calculate the
structural changes as a function of the cluster energy.
Within this method, Newton's equations of motion  are solved for
each atom within the cluster using the
Verlet algorithm \cite{Verlet}. Through this procedure, one obtains the
atomic positions and momenta as a function of time, that
are used to calculate time-averages of physical quantities
characterizing the cluster structure and dynamics.
A typical calculation consists in heating up a cluster
from  its zero temperature configuration until it transforms 
into a liquid-like cluster. To simulate this procedure
the cluster total energy is increased in a step-like manner
by scaling up the atomic velocities and therefore augmenting
the kinetic energy of the cluster.

The simulation starts by slightly perturbing
the coordinates corresponding to
the lowest-energy configuration of the cluster and setting up
the atomic momenta to a zero value, in order to eliminate
the translation of the cluster center of mass and its
rotational motion. The time step of the MD runs is 2.4 fs,
which provides total energy conservation within 0.001 \%.
For each initial condition the cluster is equilibrated during
10$^4$ time steps and the time averages are calculated
using 10$^7$ time steps in the solid and liquid regions,
but 2 times longer simulation times are used in the melting-like
transition region. This amount of averaging time is
a necessary condition to obtain converged physical quantities 
characterizing the melting process in these finite systems.

To obtain the lowest-energy structure of each cluster size
we combine simulated quenching techniques \cite{Garzon2} and 
evolutive algorithms \cite{Karo3}, which  are able to
perform global searches of the potential energy surface (PES)
in a very efficient way despite the large number of degrees
of freedom involved in the cluster structure optimization.
These methods do not only provide the lowest-energy
configuration but the distribution of isomers in the low
energy range. This optimization procedure has been successfully
used in other cluster structure 
optimizations \cite{Garzon3,Karo1,Karo2,Karo3}.

The behavior of the structural and thermal properties during
the heating process of the cluster is monitored by calculating the
temperature, heat capacity, and root-mean-square 
(rms) bond length fluctuations as a
function of the cluster total energy using the 
following expressions:

\begin{equation}
\label{tem}
T=\frac {2<E_{k}>}{(3N-6)k_{B}}
\end{equation}

\begin{equation}
\label{C}
\frac{C}{Nk_{B}}= \left[ N-N(1-\frac{2}{3N-6})<E_{k}>
<E_{k}^{-1}>\right]^{-1}
\end{equation}

\begin{equation}
\label{delta}
\delta = \frac {2}{N(N-1)} \sum^{N}_{i<j} \frac { [ < r_{ij}^{2} > -
< r_{ij} > ^{2} ] ^{1/2} } { < r_{ij} > }
\end{equation}
where $E_{k}$ is the
cluster kinetic energy, $k_{B}$ is the Boltzmann constant,
$<$...$>$ denotes a time average, and $r_{ij}$ corresponds
to the distance between atoms $i$ and $j$.
The above mathematical expressions were introduced in Refs. 
\onlinecite{Jell86,Suga} in order to calculate structural and thermal 
properties of clusters from computer simulation studies.

\section{Results and Discussion}
\subsection{Na$_{13}$}
The lowest-energy configuration of the Na$_{13}$ cluster corresponds
to an icosahedral structure shown in Fig. 1(a). 
This geometry was used to initiate
the heating up procedure through the MD method.
The cluster temperature (caloric curve) and specific heat as a function of the 
cluster total energy, calculated using Eqs. (3) and (4), 
are displayed in Figs. 2(a) and 3(a), respectively. 
The change in the slope of the caloric
curve as well as the existence of a maximum in the specific heat
are characteristics of the solid-to-liquid transition in clusters
\cite{Jell86,Suga,Jellbook}. These features are clearly seen in 
Figs. 2(a) and 3(a),
indicating a melting-like transition in the Na$_{13}$ cluster.
This transition, which occurs over a broad range of energy involving one
or more intermediate stages (such as isomerizations, coexistence of
liquidlike and solidlike phases, partial and surface melting etc.), has
been widely discussed in earlier studies of atomic clusters\cite{Melting}.
Figure 4(a) shows the rms bond-length fluctuation, $\delta$,
as a function of the cluster total energy, calculated using Eq. (5). 
It shows two abrupt variations at different energies that are in contrast
with the analog of the Lindemann criterion \cite{Lindemann} for bulk
melting, where a single abrupt increase in $\delta$ is observed.

By performing thermal quenchings from configurations generated at different
total energies during the MD simulation, the different melting
mechanisms occurring in the Na$_{13}$ cluster can be investigated.
It is found that the first abrupt change in $\delta$ at 
low energy (temperature)
is due to cluster isomerization involving only surface atoms, whereas
the second increase at higher energy (temperature) corresponds to
isomerizations where the central (bulk like) atom as well as the surface
atoms are involved. The onset of the surface and volume isomerizations occur 
at $T$=149 and 226 K, respectively, whereas the temperature corresponding
to the maximum of the specific heat (which indicates the transition
to the liquidlike state) is at $T$=260 K.
Similar isomerization mechanisms and transition temperature values
have been obtained from MC simulations of Na$_{13}$, using the same
many-body model potential \cite{Calvo2}.
A similar melting behavior has 
also been obtained for Ni$_{13}$ and Al$_{13}$ through MD simulations
using the Gupta potential \cite{Jellbook,JellIJQC}. Nevertheless, this
melting mechanism is not exclusive of metal clusters interacting by a
many-body Gupta potential since both surface and volume isomerizations
have also been obtained for a 13-atom cluster modeled by a pair-wise
additive potential with a soft repulsive core interaction \cite{Rey}.

\subsection{Na$_{20}$}

The double icosahedron with a capped atom over the central pentagonal
ring displayed in Fig. 1(b) corresponds to 
the lowest-energy structure of Na$_{20}$.
By heating up this cluster we obtain the caloric curve and specific
heat displayed in Figs. 2(b) and 3(b), 
and the rms bond length fluctuation shown
in Fig. 4(b). 
In addition to the slight change in the slope of the caloric curve
at low energies, the specific heat shows a shoulder at low energy
(temperature), before it reaches its maximum at higher energy (temperature).
A very abrupt increase in the $\delta $ value is obtained at low energy that
corresponds, according to the caloric curve, to a cluster temperature
of 57 K. A second abrupt increase in $\delta $ at around 
157 K is also obtained
after further heating of Na$_{20}$. 

The microscopic features 
characterizing this melting behavior can be extracted by performing
periodic quenchings using the cluster configurations along the MD
run at a given cluster energy. The analysis of the quenched cluster
structures indicate that the first abrupt increase in $\delta $, that
is in correspondence with the shoulder in the specific heat at low
energies, is related with isomerization transitions where the
extra atom incorporates into the cluster surface, generating
a fluctuating six-atom ring in the central region of the cluster
structure. This isomerization is equivalent to a surface 
reconstruction of the icosahedral surface without diffusion.
The second jump in $\delta $ is associated to a surface melting
stage where the double icosahedron transforms into a more compact
structure containing two internal atoms. At higher energies (temperatures)
complete melting is observed, characterized by diffusive
motion where the $\delta $ value levels off.
The two abrupt increases in $\delta $ obtained for Na$_{20}$ are equivalent
to those described for Na$_{13}$, except that in the larger cluster
the first increase corresponds to an isomerization due to the
outer atom lying over the cluster surface.
These phenomena occurring in Na$_{13}$ and Na$_{20}$ before they
fully melt, showing atomic diffusion, 
can be denominated as a \it premelting
\rm stage. 

The premelting phenomenon was first obtained in MD simulations
of Ni$_{14}$ and Ni$_{20}$ \cite{JellNi14,GarNi20,Guvenc} and some
Be$_{N}$, $N$=9,11,12,14, clusters \cite{Roman}. Later,
other metal clusters with magic and non-magic number sizes
also reveled a premelting stage (see, for example, 
Refs. \onlinecite{Calvo2,Jellbook}).
The two-step melting transition described above was also obtained
using the q-jumping MC method and the Gupta potential \cite{Calvo1,Calvo2}.
A good agreement was found in the premelting and melting temperatures
calculated with this approach \cite{Calvo1,Calvo2}.
However, higher temperature values, for the two
transitions, were found by
first-principles orbital-free MD on Na$_{20}$ \cite{Agua1}, which might be
due to the much shorter simulation time used in such calculation.

\subsection{Na$_{55}$}
Figure 1(c) shows the lowest-energy configuration of 
Na$_{55}$ corresponding to the
two-shell Mackay icosahedron. The caloric curve and specific heat
obtained by heating up this structure are displayed in Figs. 2(c) and
3(c), respectively.
Figure 4(c) shows the rms-bond length fluctuation where an abrupt 
increase is observed at a total energy that, according to the
caloric curve, indicates a melting temperature of 151 K, whereas
the maximum in the specific heat is obtained at a slightly higher
temperature of 166 K. In contrast to Na$_{13}$ and Na$_{20}$,
the Na$_{55}$ cluster shows a single abrupt change in $\delta$.
However, by visualizing the dynamical trajectories of each atom
in the cluster at the temperatures within the transition
region, we found that the melting process develops in several
stages. In the first one, the most external layer fluctuates
between an incomplete and complete icosahedral surface by
expelling and receiving atoms back and forth. At higher energies
an exchange of atoms between the intermediate and most
external layers is obseved, and with a further increase
in energy, fully melting is found, where the central atom also
contributes to the cluster diffusive motion. This complex melting
mechanism has been recently studied in Al$_{55}$ and other 
aluminum clusters  by introducing
dynamical degrees of freedom \cite{JellJCP}.

Similar melting stages have also been obtained by
MC simulation of Na$_{55}$ \cite{Calvo1,Calvo2}. However, in
those calculations a slightly higher melting temperature of
175 K was reported. The first-principles orbital-free 
MD simulation of Na$_{55}$
also reported \cite{Agua2} a melting transition at 190 K.
Again, the smaller melting temperature
obtained in the present work might be due to the much longer simulation
times we have used in our MD simulations as compared to those
used in Ref.\onlinecite{Agua2}. Nevertheless, none of the
melting temperatures calculated by us and other authors
\cite{Calvo1,Calvo2,Agua2} are in agreement with the experimental
value (320 K) reported for Na$_{55}$ \cite{Haber3}.
This discrepancy between theory and experiment has not yet
been solved since it would require a more
detailed modeling of the energy landscape of Na$_{55}$, that
includes not only information on the basins of attraction of the
equilibrium structures but also
on the topology around saddle points connecting the lowest-energy
minima. Additionally, in a more detailed description of the potential energy
surface it would be possible that the global minimum for Na$_{55}$ 
does not correspond to the icosahedral structure but to an unknown
special geometry with larger thermal stabiltity against melting.
At present, it represents a theoretical challenge to characterize
the potential energy surface of systems with such number
of degrees of fredom using first-principles methods, however
intense efforts are currently being performed to solve this problem.
On the other hand, further experimental work is expected in the near 
future that confirms the relatively high value of the melting temperature 
of the Na$_{55}$ cluster \cite{Haber3}. 

\subsection{Na$_{135}$, Na$_{142}$, and Na$_{147}$}
The global minima of the larger sodium clusters investigated in this work
correspond to icosahedral structures. The three-layer
Mackay icosahedron shown in Fig. 1(f), 
is the lowest-energy isomer of Na$_{147}$.
The lowest-energy structures of the Na$_{135}$, Na$_{142}$
are incomplete icosahedra obtained by removing 12 and 5 
vertex atoms, respectively, from the 147 Mackay icosahedron
(see Figs. 1(d) and 1(e), respectively).
Despite the existence of an incomplete surface layer in
the lowest-energy structure of Na$_{135}$, Na$_{142}$, they
show caloric curves, Figs. 2(d-e), and 
rms bond-length fluctuations, Figs. 4(d-e),
very similar to those obtained by heating up the complete 
icosahedron structure of the 147-atom cluster, Figs. 2(f) and 4(f).
The calculated melting temperatures obtained using the
Lindemann criterion ($\delta$ $\sim$ 0.15) for 
Na$_{135}$, Na$_{142}$, and Na$_{147}$ are 135 K, 190 K, and
171 K, respectively. These values are smaller
than those obtained from the maximum of the specific heat
(see Table I). By visualizing the atomic
coordinates as a function of time at the energies where
the $\delta$ values change abruptly, it is observed that
the melting is initiated at the cluster surface. For the
three sizes investigated, the atomic mobility increases with
temperature starting from
the most external layer and propagating into the internal
layers. This stage, known as surface melting \cite{Guvenc},
precedes the complete cluster melting characterized by the
diffusive motion of all the atoms in the cluster which is
observed at temperatures where the specific heat is maximum.

Our calculated melting temperatures for these larger sodium clusters are
lower than those obtained by other authors using the MC method 
with the same potential \cite{Calvo1,Calvo2} and the orbital
free MD simulations \cite{Agua2}. This difference, as in the
smaller cluster sizes, we assume is due to the much longer simulation
times we have used in our MD calculations. On the other hand,
although our results show that the largest melting temperature
corresponds to the Na$_{142}$ cluster, in agreement with the
experimental data \cite{Haber2,Haber3}, the absolute values
of our calculated melting temperatures are about 30 $\%$ lower
than the experimental values \cite{Haber2,Haber3}.
These results indicate that the many-body Gupta potential,
which does not include electronic degrees of freedom, 
only provide a qualitative description of the melting of 
sodium clusters.

\section{Summary}
The melting-like transition of Na$_N$, N = 13, 20, 55, 135, 142, and 147,
has been investigated through microcanonical MD simulations
using a phenomenological many-body Gupta potential.
The solid-to-liquid transition was studied by calculating
caloric curves, rms bond-length-fluctuations and specific heats.
The indicators of the cluster melting correspond to
changes in the slope of the caloric curve, abrupt increases in
the  $\delta$ values, and the existence of maxima in the specific heats.
Table I shows the melting temperatures calculated for all cluster
sizes using those criteria. The main features coming out from
these data are: (\it i \rm) The melting temperatures calculated from the
maxima of the specific heats are systematically higher than
the values obtained using the Lindemann criterion.
(\it ii \rm) There is
an irregular variation in the melting temperatures as a function
of the cluster size, the highest value being the one corresponding
to the Na$_{142}$ cluster. These results are in qualitative agreement with
the experimental data \cite{Haber2,Haber3}. 
(\it iii \rm) The calculated melting
temperature for the Na$_{55}$ cluster is about 40 $\%$ lower than
the reported experimental value \cite{Haber3}.
(\it iv \rm) The melting transition in sodium clusters is a complex
phenomenon that involves several stages in which the system
undergoes different structural changes (isomerization, premelting
and surface melting) before it shows a diffusive regime characteristic
of the liquidlike phase.

A comparison of the present results with those obtained using
the same many-body potential and the MC method \cite{Calvo1,Calvo2},
and with those generated from orbital-free MD simulations
\cite{Agua1,Agua2}, indicate
that in general, the melting temperatures calculated by
heating up the lowest-energy isomer, are lower when much longer
simulation time is employed. In this work, the simulation time
was extended up to the nanosecond time regime where it is very likely that
the time-averages of the physical quantities that characterize
the cluster melting might be much better converged.

In order to obtain a better (quantitative) agreement with the
experimental results on the melting of sodium clusters it would be
necessary to either extend the simulation time in the first-principles
MD calculations or to design a many-body potential that describes
with a higher level of approximation the complex topology of the
potential energy landscape of sodium clusters. Work in both
directions is currently under progress.

\begin{acknowledgements}
This work was supported by
DGSCA-UNAM Supercomputing Center. JARN acknowledges
a graduate fellowship from DGEP-UNAM.
\end{acknowledgements}

\begin{table}
\caption{Binding energies (BE) in eV/atom and melting temperatures in K
calculated from the temperature value
at the maximum of the specific heat and using the Lindemann
criterion ($\delta$ $\sim$ 0.15). For $N$ = 13 and 20 there are two values 
due to the existence of two-stage 
(premelting and melting) transitions.} 
\begin{tabular}{ccccc}
$N$ & BE & Maximum in $C$ & Lindemann criterion & Exp.$^a$ \\
\hline
13 & 0.684 & 260  & 149, 226 &  \\
20 & 0.734 & 220  & 57, 157  &  \\
55 & 0.855 & 166  & 151  & 320 \\
135 & 0.929 &  181  & 135 & 250 \\
142 & 0.933 & 189  & 190 & 285  \\
147 & 0.935 & 180  & 171  & 272 \\
\end{tabular}
$^a$Refs. \onlinecite{Haber2,Haber3}.
\label{melting}
\end{table}

\begin{figure}
\caption{Lowest-energy structures of
Na$_{N}$, $N$ = 13 (a); 20 (b); 55 (c); 135 (d); 142 (e);
and 147 (f); clusters. The cluster geometries correspond to
Mackay icosahedra for $N$ = 13, 55, and 147; a capped double
icosahedron for $N$ = 20; and incomplete three-layer Mackay
icosahedra for $N$ = 135 and 142.}
\label{isomeros}
\end{figure}

\begin{figure}
\caption{Caloric curves of 
Na$_{N}$, $N$ = 13 (a); 20 (b); 55 (c); 135 (d); 142 (e);
and 147 (f); clusters. The cluster energy is calculated taking
as reference the value of the binding energy of the most-stable 
(lowest-energy)
configuration given in Table I.}
\label{caloric}
\end{figure}

\begin{figure}
\caption{Specific heats of Na$_{N}$, $N$ = 13 (a); 20 (b);
and 55 (c); clusters. The cluster energy is calculated taking
as reference the value of the binding energy of the most-stable
(lowest-energy)
configuration given in Table I.}
\label{heats}
\end{figure}

\begin{figure}
\caption{RMS bond-length fluctuations of Na$_{N}$,
$N$ = 13 (a); 20 (b); 55 (c); 135 (d); 142 (e);
and 147 (f); clusters.
The cluster energy is calculated taking
as reference the value of the binding energy of the most-stable
(lowest-energy)
configuration given in Table I.}
\label{rms}
\end{figure}

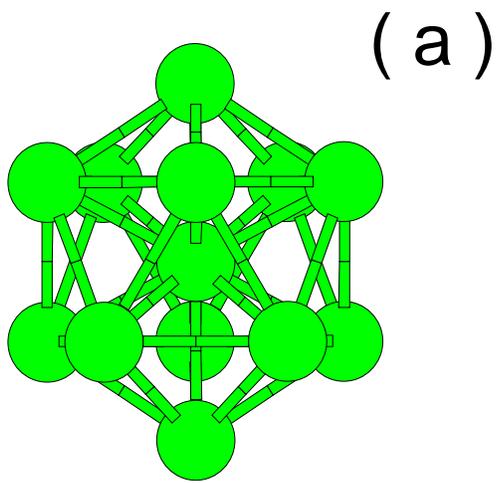
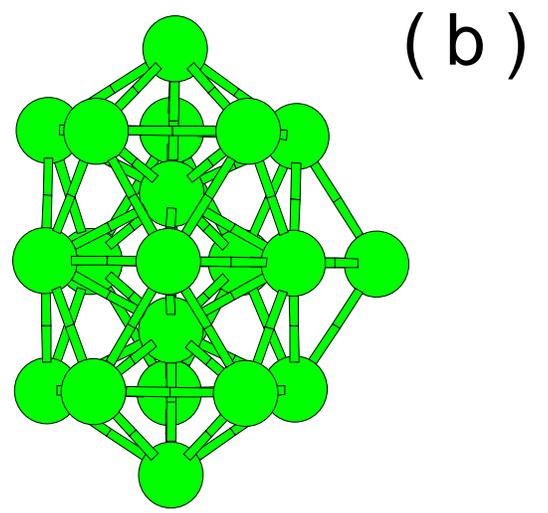
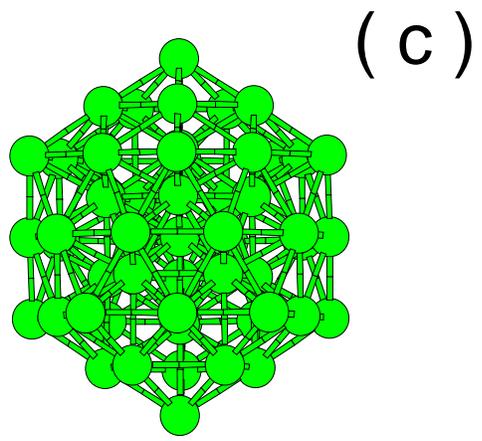
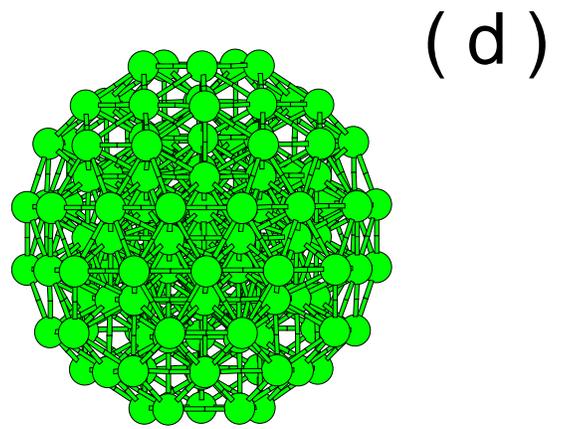
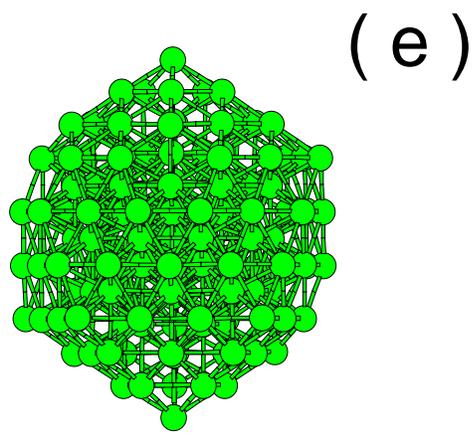
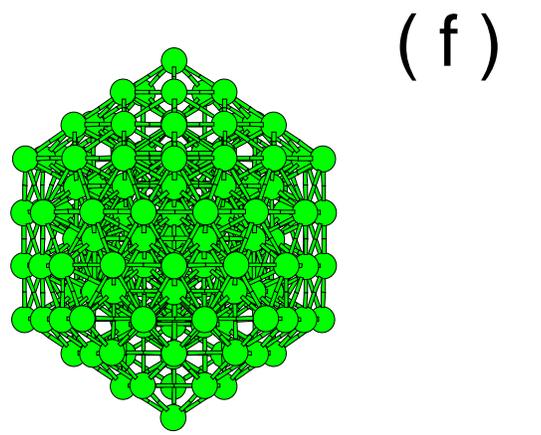
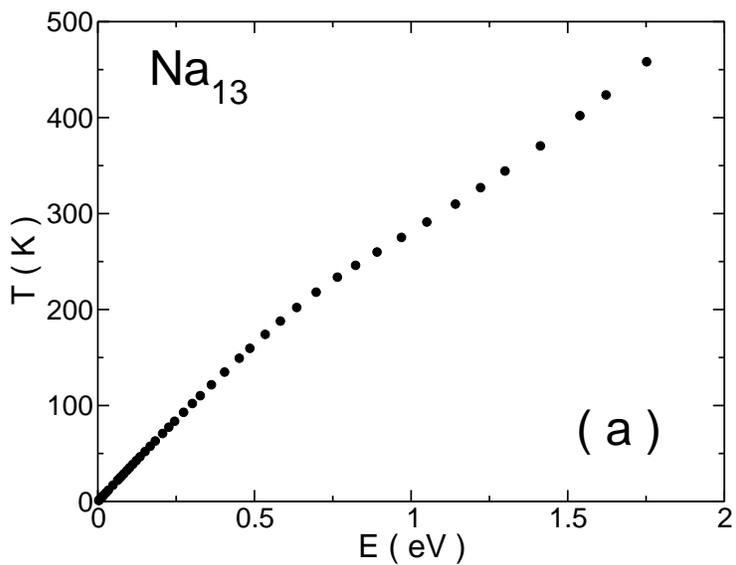
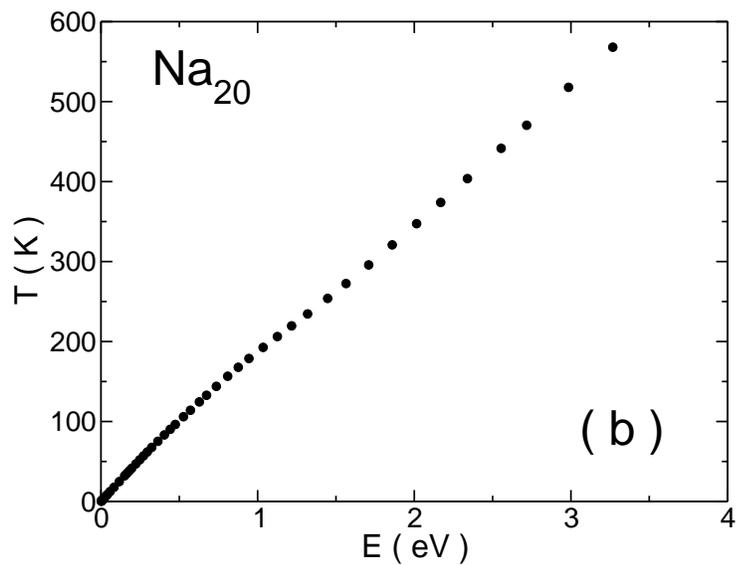
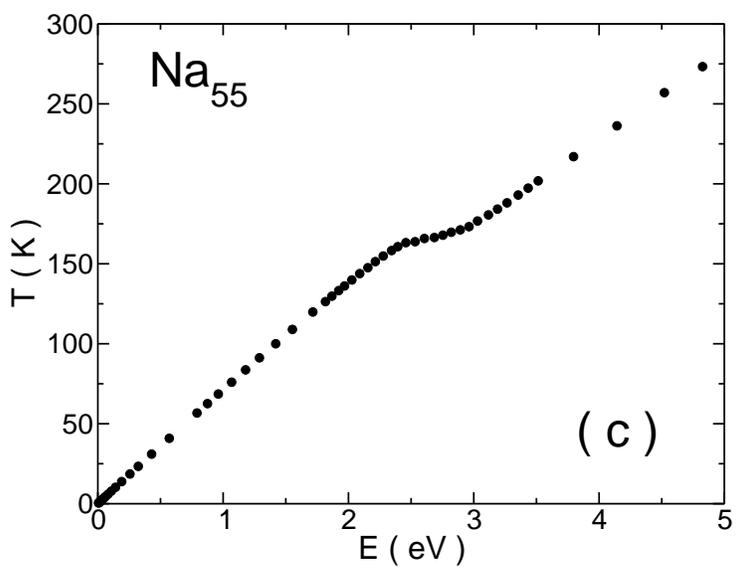
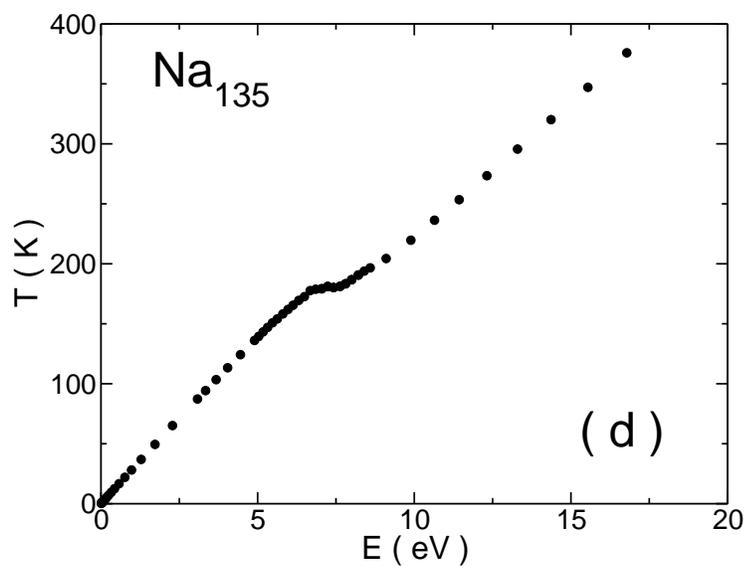
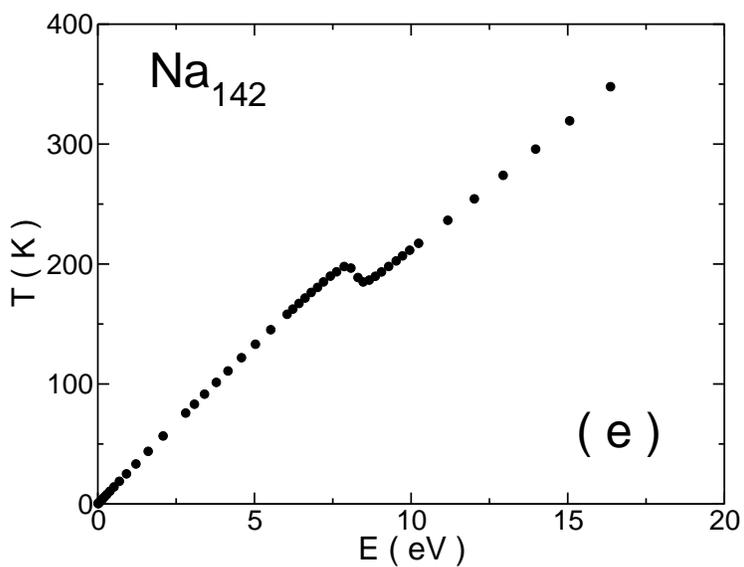
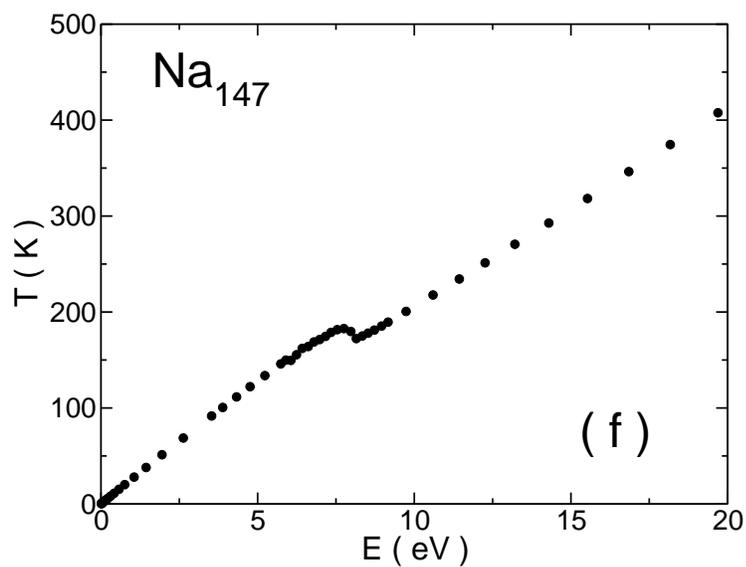
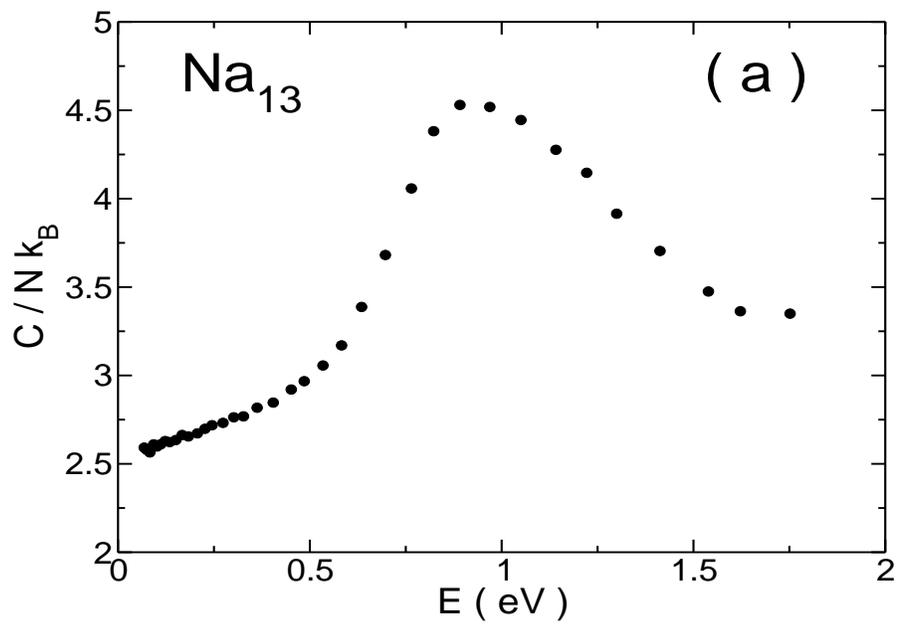
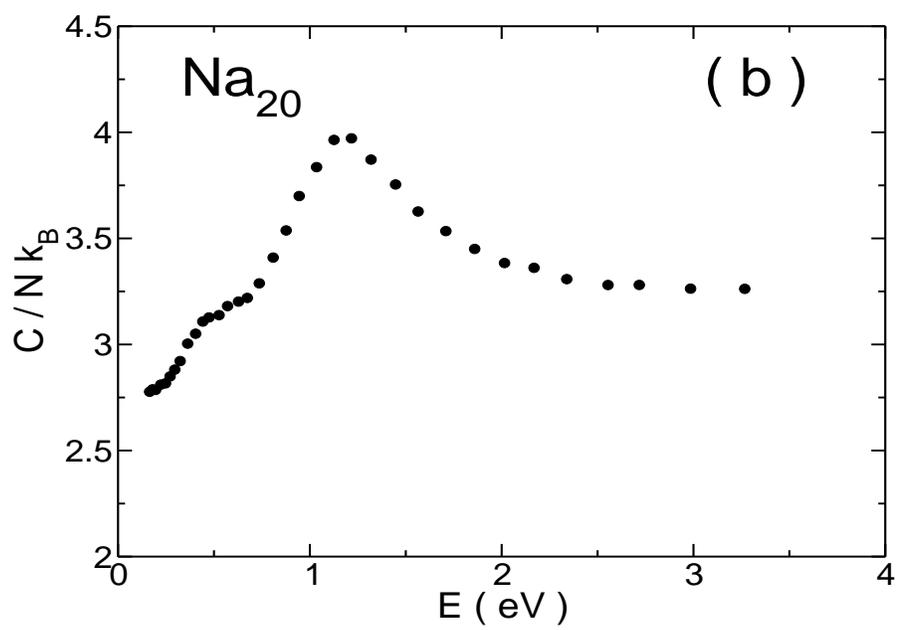
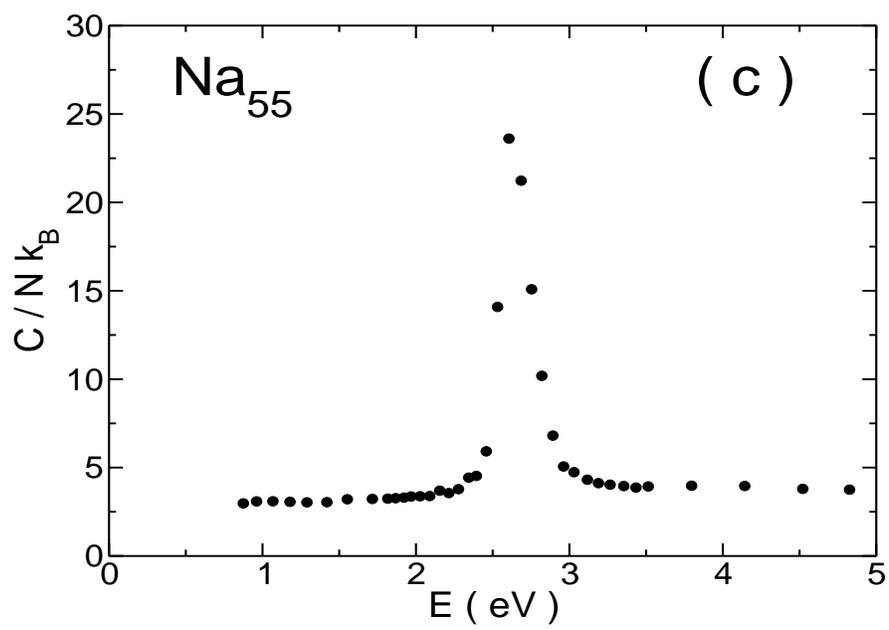
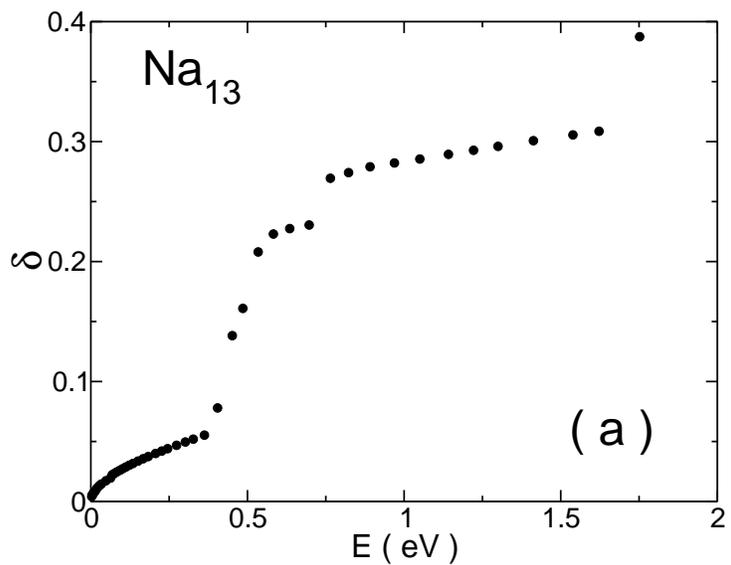
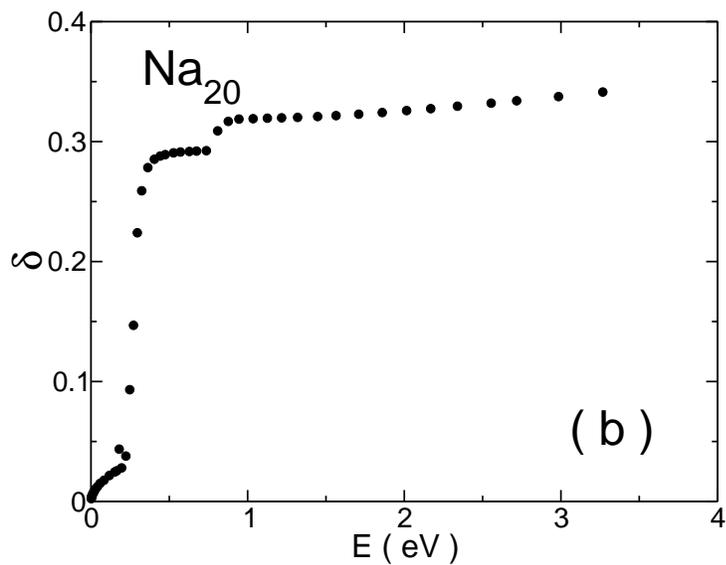
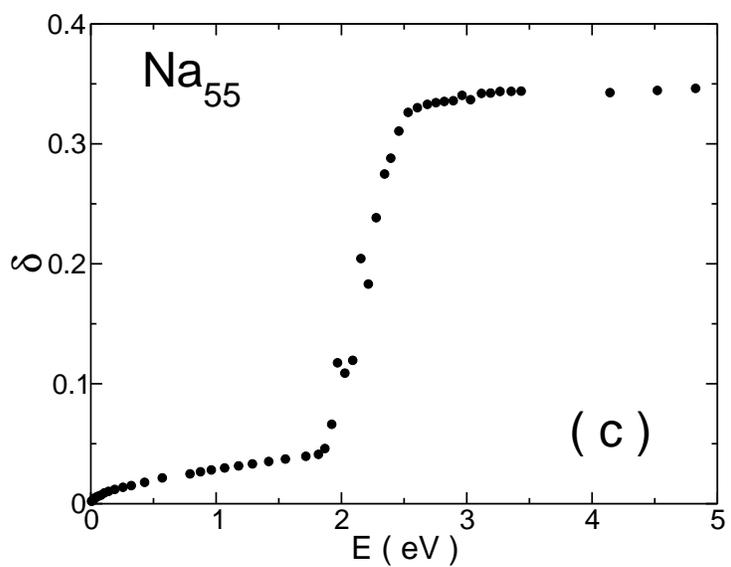
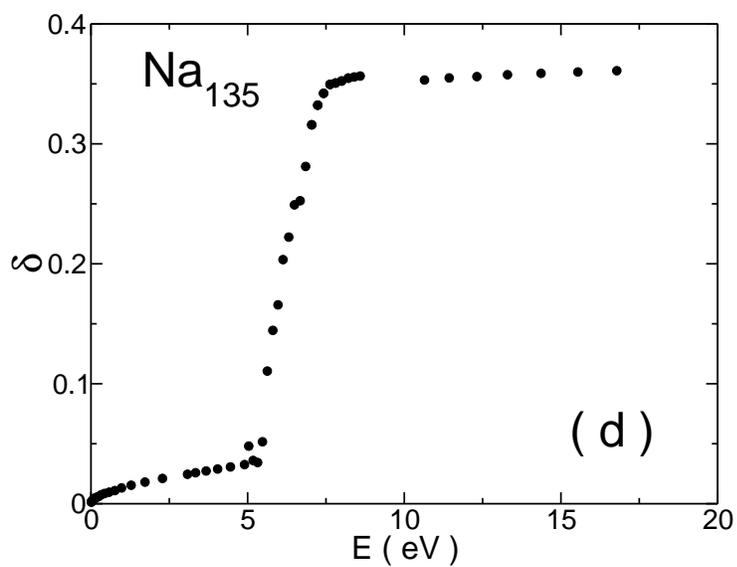
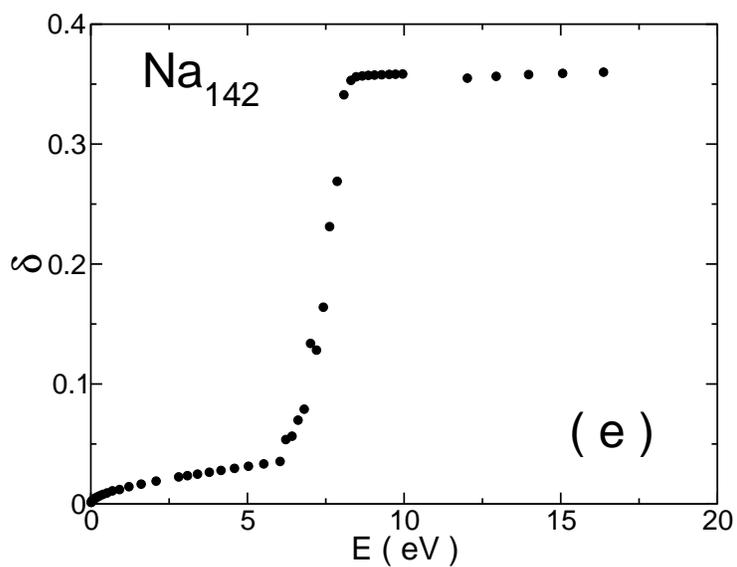
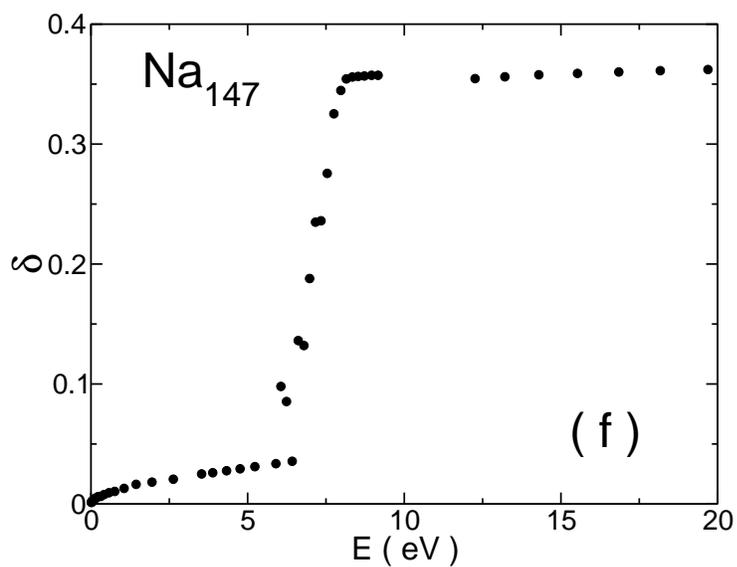
\end{document}